\documentclass[twocolumn,prb,showpacs,byrevtex,superbib]{revtex4}
\usepackage{amsmath}
\usepackage{graphicx}
\usepackage{epsfig}
\usepackage[bf]{subfigure}
\usepackage[scaled=.92]{helvet}
\usepackage{courier}
\usepackage{mathptmx}

\input{tcilatex}
\DeclareMathAlphabet{\mathsf}{OT1}{cmss}{bx}{n}

\begin{document}

%TCIMACRO{
%\TeXButton{title}{\title
%[Ray Stability]{Ray stability in weakly range-dependent sound channels}}}%
%BeginExpansion
\title
[Ray Stability]{\Large{Ray stability in weakly range-dependent sound channels}}%
%EndExpansion

%TCIMACRO{
%\TeXButton{author}{\author{F. J. Beron-Vera$^{\textrm{a})}$\footnotetext{$^{\textrm{a})}$Author to whom correspondence should be addressed. Electronic mail: \texttt{fberon@rsmas.miami.edu}}}}}%
%BeginExpansion
\author{\textsf{F. J. Beron-Vera}$^{\textrm{a})}$\footnotetext{$^{\textrm{a})}$Author to whom correspondence should be addressed. Electronic mail: \texttt{fberon@rsmas.miami.edu}}}%
%EndExpansion

%TCIMACRO{
%\TeXButton{affiliation}{\affiliation{RSMAS/AMP, University of Miami, Miami, FL 33149}}}%
%BeginExpansion
\affiliation{RSMAS/AMP, University of Miami, Miami, FL 33149}%
%EndExpansion

%TCIMACRO{\TeXButton{author}{\author{M. G. Brown}}}%
%BeginExpansion
\author{\textsf{M. G. Brown}}%
%EndExpansion

%TCIMACRO{
%\TeXButton{affiliation}{\affiliation{RSMAS/AMP, University of Miami, Miami, FL 33149}}}%
%BeginExpansion
\affiliation{RSMAS/AMP, University of Miami, Miami, FL 33149}%
%EndExpansion

%TCIMACRO{\TeXButton{keywords}{\keywords{Ray stability}}}%
%BeginExpansion
%EndExpansion

%TCIMACRO{\TeXButton{pacs}{\pacs{43.30.Cq, 43.30.Ft, 43.30.Pc}}}%
%BeginExpansion
%EndExpansion

%TCIMACRO{\TeXButton{Begin abstract}{\begin{abstract}}}%
%BeginExpansion
\begin{abstract}%
%EndExpansion

Ray stability is investigated in environments consisting of a
range-independent background sound-speed profile on which a range-dependent
perturbation is superimposed. Theoretical arguments suggest and numerical
results confirm that ray stability is strongly influenced by the background
sound speed profile. Ray instability is shown to increase with increasing
magnitude of $\alpha (I)=(I/\omega )\mathrm{d}\omega /\mathrm{d}I,$ where $%
2\pi /\omega (I)$ is the range of a ray double loop and $I$ is the ray
action variable. This behavior is illustrated using internal-wave-induced
scattering in deep ocean environments and rough surface scattering in upward
refracting environments.

%TCIMACRO{\TeXButton{End abstract}{\end{abstract}}}%
%BeginExpansion
\end{abstract}%
%EndExpansion

%TCIMACRO{\TeXButton{volumeyear}{\volumeyear{2002}}}%
%BeginExpansion
\volumeyear{2002}%
%EndExpansion

%TCIMACRO{\TeXButton{volumenumber}{\volumenumber{205}}}%
%BeginExpansion
\volumenumber{205}%
%EndExpansion

%TCIMACRO{\TeXButton{issuenumber}{\issuenumber{5}}}%
%BeginExpansion
\issuenumber{5}%
%EndExpansion

%TCIMACRO{\TeXButton{date}{\date[Dated: ]{\today}}}%
%BeginExpansion
\date[Dated: ]{\today}%
%EndExpansion

%TCIMACRO{\TeXButton{startpage}{\startpage{1}}}%
%BeginExpansion
\startpage{1}%
%EndExpansion

%TCIMACRO{\TeXButton{endpage}{\endpage{102}}}%
%BeginExpansion
\endpage{102}%
%EndExpansion

%TCIMACRO{\TeXButton{eid}{\eid{identifier}}}%
%BeginExpansion
\eid{identifier}%
%EndExpansion

%TCIMACRO{\TeXButton{maketitle}{\maketitle}}%
%BeginExpansion
\maketitle%
%EndExpansion

\section{\textsf{Introduction}}

Measurements made during the Slice89 propagation experiment \cite%
{Duda-et-al-92}, made in the eastern North Pacific, showed a clear contrast
between highly structured steep ray arrivals and relatively unstructured
near-axial flat ray arrivals. Measurements made during the AET experiment %
\cite{Worcester-et-al-99,Colosi-et-al-99} in a similar environment provided
further evidence of the same behavior. Motivated by these observations,
several authors \cite%
{Duda-Bowlin-94,Simmen-Flatte-Yu-Wang-97,Smirnov-Virovlyansky-Zaslavsky-01}
have investigated ray sensitivity to environmental parameters. The work
described in this paper continues the same line of investigation.

Like the earlier work we focus on ray path stability in physical space or
phase space. The extension to travel time stability is not considered in
this paper. Clearly, however, travel time stability must be addressed to
fully understand the Slice89 and AET measurements. In spite of this
limitation, and limitations of the ray approximation, our results are in
agreement---qualitatively, at least---with the Slice89 and AET observations.

In Sec. \ref{TheoBack} we provide the theoretical background for the work
that follows. Most of this material builds on the action--angle form of the
ray equations. A trivial observation that follows from the angle--action
formalism is that ray sensitivity to the background sound speed structure is
controlled by the function $\omega (I),$ where $2\pi /\omega $ is the range
of a ray double loop and $I$ is the action variable. A heuristic argument
suggests that $\mathrm{d}\omega /\mathrm{d}I$ should be closely linked to
ray stability. In Sec. \ref{NumSim} we present numerical simulations that
are chosen to demonstrate the importance of the background sound speed
structure on ray stability. Simulations are shown for both
internal-wave-induced scattering in deep ocean environments and rough
surface scattering in upward-refracting environments. The latter are
included to demostrate the generality of the arguments presented. These
simulations strongly suggest that ray stability is controlled by the
magnitude of the nondimensional quantity
\begin{equation}
\alpha (I)=\frac{I}{\omega }\frac{\mathrm{d}\omega }{\mathrm{d}I}.
\label{Alpha}
\end{equation}%
In Sec. \ref{ShearIns} we explain the mechanism by which $\left| \alpha
\right| $ controls ray stability. In Sec. \ref{SumDis} we summarize our
results and briefly discuss: (i) the relationship between our work and
earlier investigations; (ii) timefront stability; (iii) the dynamical
systems viewpoint; and (iv) the extension to background sound speed
structures with range-dependence. An appendix is reserved for some
mathematical details.

\section{\textsf{Theoretical background}\label{TheoBack}}

\subsection{\textsf{One-way ray equations}}

We consider underwater sound propagation in a two-dimensional waveguide with
Cartesian coordinates $z$ (upward) and $r$ (along-waveguide). One-way ray
trajectories satisfy the\textit{\ \textbf{canonical Hamilton's equations}\ }%
(cf. e.g. Ref.
%TCIMACRO{\TeXButton{Brown-etal-03}{\onlinecite{Brown-etal-03}} }%
%BeginExpansion
\onlinecite{Brown-etal-03}
%EndExpansion
and references therein),%
\begin{equation}
\frac{\mathrm{d}p}{\mathrm{d}r}=-\frac{\partial h}{\partial z},\quad \frac{%
\mathrm{d}z}{\mathrm{d}r}=\frac{\partial h}{\partial p},  \label{RayEqn}
\end{equation}%
with Hamiltonian%
\[
h(p,z;r)=-\sqrt{c^{-2}-p^{2}}.
\]%
Here, $c(z;r)$ is the sound speed; $p$ is the\textit{\ }vertical slowness%
\textit{\ }which is understood as the momentum, conjugate to the generalized
coordinate $z$; and $r$ is the independent (time-like) variable. The
vertical slowness and the sound speed are related through $p\,c=\sin \varphi
,$ where $\varphi $ is the angle that a ray makes with the horizontal.

\subsection{\textsf{Near-integrability under small waveguide perturbations}}

Assume that the sound speed can be split as into a background
(range-independent) part, $C(z),$ and a small range-dependent perturbation, $%
\delta c(z;r).$ Then, to lowest-order in $\delta c/c,$ the Hamiltonian takes
the additive form%
\[
h=H(p,z)+\delta h(p,z;r).
\]%
Introduce now the\textit{\ }Poincar\'{e} action\textit{\ }%
\begin{equation}
I=\frac{1}{2\pi }\oint \mathrm{d}z\,p=\frac{1}{\pi }\int_{z_{-}}^{z_{+}}%
\mathrm{d}z\,\sqrt{C^{-2}-H^{2}},  \label{I}
\end{equation}%
where $z_{\pm }$ denotes the vertical coordinate of the upper ($+$) and
lower ($-$) turning points of a ray double loop, and consider the canonical
transformation into\textit{\ \textbf{action--\textit{angle} variables}\ }$%
(p,z)\mapsto (I,\vartheta )$ defined by%
\[
p=\frac{\partial G}{\partial z},\;\vartheta =\frac{\partial G}{\partial I}%
,\;G(z,I)=\int_{z_{-}}^{z}\mathrm{d}\xi \,\sqrt{C^{-2}(\xi )-H^{2}}.
\]

According to the above transformation,
\[
h(p,z;r)\mapsto \bar{H}(I)+\delta \bar{h}(I,\vartheta ;r)
\]%
and the ray equations (\ref{RayEqn}) take the form%
\begin{equation}
\frac{\mathrm{d}I}{\mathrm{d}r}=-\frac{\partial }{\partial \vartheta }\delta
\bar{h},\quad \frac{\mathrm{d}\vartheta }{\mathrm{d}r}=\omega +\frac{%
\partial }{\partial I}\delta \bar{h},  \label{NeaInt}
\end{equation}%
where%
\begin{equation}
\omega (I)=\frac{\mathrm{d}\bar{H}}{\mathrm{d}I}.  \label{omega}
\end{equation}%
Set (\ref{NeaInt}) constitutes a\textit{\ \textbf{near-integrable
Hamiltonian system}\ }for the reasons explained next.

In the limit $\delta \bar{h}\rightarrow 0,$ the ray equations (\ref{NeaInt}%
), which have one degree of freedom, are autonomous and the corresponding
Hamiltonian, $\bar{H},$ is an integral of motion that constrains the
dynamics. As a consequence, the equations are integrable through quadratures
and the motion is \textit{periodic} with (spatial) frequency $\omega $.
Namely $I=I_{0}$ and $\vartheta =\vartheta _{0}+\omega r$ $\func{mod}2\pi ,$
where $I_{0}$ and $\vartheta _{0}$ are constants. Every solution curve is
thus a line that winds around an invariant one-dimensional torus, whose
representation in $(p,z)$-space is the closed curve given by the isoline $H=%
\bar{H}(I_{0}).$ Notice that $C^{-1}\cos \varphi =H;$ consequently, each
torus can be uniquely identified by the ray\textit{\ }axial angle\textit{\ }%
defined by $\varphi _{a}(p,z)=\arccos C_{a}H$, where $C_{a}$ is the
background sound speed at the sound channel axis.

With nonzero $\delta \bar{h}$, the Hamiltonian, $\bar{H}+\delta \bar{h},$ is
no longer an invariant (the equations are nonautonomous) and the system may
be\ \textit{sensitive to initial conditions}, leading to\textit{\ chaotic
motion }in phase space. The distinction between regular and chaotic ray
trajectories is commonly quantified by the\textit{\ \textbf{Lyapunov exponent%
}},\ formally defined by%
\begin{equation}
\nu _{\infty }=\lim_{r\rightarrow \infty }\lim_{d_{0}\rightarrow 0}\frac{1}{r%
}\ln \frac{d}{d_{0}},  \label{ForLyaExp}
\end{equation}%
where $d(r),$ such that $d(0)=d_{0},$ is a suitably chosen measure of the
separation between neighboring ray trajectories in phase space. For regular
trajectories, $d\sim r^{a}$ as $r\rightarrow \infty $ with $a>0$ and, hence,
$\nu _{\infty }=0.$ For chaotic trajectories, instead, $d\sim \exp \nu
_{\infty }r$; in this case, $\nu _{\infty }^{-1}$, the average $\mathrm{e}$%
-folding range, is regarded as the\textit{\ \textbf{predictability\ horizon}.%
}

\subsection{\textsf{Variational equations}}

Although any norm of $(\delta p,\delta z),$ a perturbation to a trajectory $%
(p,z)$, could be used to define a distance in phase space, this is not a
trivial task because $z$ and $p$ do not have the same dimensions, which
complicates the evaluation of (\ref{ForLyaExp}). The following procedure
eliminates this problem.

The\textit{\ \textbf{variational equations}}, which follow from the ray
equations (\ref{RayEqn}), are%
\begin{equation}
\frac{\mathrm{d}\mathsf{Q}}{\mathrm{d}r}=\mathsf{JQ},\quad \mathsf{Q}(0)=%
\mathsf{I},  \label{RayVar}
\end{equation}%
where%
%TCIMACRO{\TeXButton{llaves-ini}{\renewcommand{\arraystretch}{1.5}}}%
%BeginExpansion
\renewcommand{\arraystretch}{1.5}%
%EndExpansion
\[
\mathsf{J}=\left[
\begin{array}{cc}
-\dfrac{\partial ^{2}h}{\partial p\partial z} & -\dfrac{\partial ^{2}h}{%
\partial z^{2}} \\
\dfrac{\partial ^{2}h}{\partial p^{2}} & \dfrac{\partial ^{2}h}{\partial
p\partial z}%
\end{array}%
\right] ,\quad \mathsf{Q}=\left[
\begin{array}{cc}
\dfrac{\partial p}{\partial p_{0}} & \dfrac{\partial p}{\partial z_{0}} \\
\dfrac{\partial z}{\partial p_{0}} & \dfrac{\partial z}{\partial z_{0}}%
\end{array}%
\right] ,
\]%
%TCIMACRO{\TeXButton{llaves-fin}{\renewcommand{\arraystretch}{1.00}}}%
%BeginExpansion
\renewcommand{\arraystretch}{1.00}%
%EndExpansion
with $p_{0}=p(0)$ and $z_{0}=z(0),$ and $\mathsf{I}$ is the identity matrix.
Here, $\mathsf{J}(r)$ and $\mathsf{Q}(r)$ are the Jacobian matrices of the
Hamiltonian vector field and associated flow, respectively; the last is
usually referred to as the\textit{\ \textbf{stability matrix}.} Notice that $%
(\delta p,\delta z)^{\mathrm{T}}=\mathsf{Q}(\delta p_{0},\delta z_{0})^{%
\mathrm{T}}$ at the lowest-order in $(\delta p_{0},\delta z_{0})$. Let now $%
\nu ^{\mathsf{Q}}(r)$ be the largest of the two eigenvalues of $\mathsf{Q},$
and consider the definition (cf. e.g. Ref.
%TCIMACRO{\TeXButton{Parker-Chua-89}{\onlinecite{Parker-Chua-89}}}%
%BeginExpansion
\onlinecite{Parker-Chua-89}%
%EndExpansion
)%
\begin{equation}
\nu _{\infty }=\lim\limits_{r\rightarrow \infty }\dfrac{1}{r}\ln |\nu ^{%
\mathsf{Q}}|\text{\textit{.}}  \label{LyaExp}
\end{equation}%
If the limit in (\ref{LyaExp}) exists, and is not nil, then $(\delta
p,\delta z)\sim \exp \nu _{\infty }r$ as $r\rightarrow \infty $ (nearby
trajectories diverge exponentially in range) and, hence, (\ref{LyaExp}) can
be taken as a suitable definition of the Lyapunov exponent. The equivalence
between (\ref{ForLyaExp}) and (\ref{LyaExp}) can be understood by noting
that the variational equations describe the evolution of an infinitesimal
circle of initial conditions surrounding a specified ray initial condition.
The circle gets deformed into an ellipse whose area equals that of the
initial circle according to Liouville's theorem (cf. e.g. Refs.
%TCIMACRO{\TeXButton{Arnold-89,Tabor-89}{\onlinecite{Arnold-89,Tabor-89}}}%
%BeginExpansion
\onlinecite{Arnold-89,Tabor-89}%
%EndExpansion
). The eigenvectors of \textsf{Q\ }define the orientation of the ellipse.
The largest eigenvalue is a measure of the length of the semimajor axis and,
hence, a suitable choice of $d.$

A simple but very important observation follows from the action--angle
formalism. Dependence of the ray and variational equations on the background
sound speed structure enters only through the function $\omega (I)$. The
action--angle form of the variational equations for the perturbed system
strongly suggests that ray stability and $\mathrm{d}\omega /\mathrm{d}I$ are
closely linked$.$ The mechanism through which $\mathrm{d}\omega /\mathrm{d}I$
influences ray stability can be seen from the action--angle form of the ray
variational equations,%
\begin{eqnarray*}
\frac{\mathrm{d}}{\mathrm{d}r}\delta I &=&-\dfrac{\partial ^{2}\delta \bar{h}%
}{\partial I\partial \vartheta }\delta I-\dfrac{\partial ^{2}\delta \bar{h}}{%
\partial \vartheta ^{2}}\delta \vartheta , \\
\frac{\mathrm{d}}{\mathrm{d}r}\delta \vartheta  &=&\dfrac{\mathrm{d}\omega }{%
\mathrm{d}I}\delta I+\dfrac{\partial ^{2}\delta \bar{h}}{\partial I^{2}}%
\delta I+\dfrac{\partial ^{2}\delta \bar{h}}{\partial I\partial \vartheta }%
\delta \vartheta .
\end{eqnarray*}%
If one assumes that the second derivatives of $\delta \bar{h}$ are zero-mean
random variables, then when $\mathrm{d}\omega /\mathrm{d}I=0$ these terms
should lead to slow (power-law) growth of $\delta \vartheta $ and $\delta I.$
But if $\left| \mathrm{d}\omega /\mathrm{d}I\right| $ is large, this term
will cause $\left| \delta \vartheta \right| $ to rapidly grow for any
nonzero $\left| \delta I\right| $. The perturbation terms will then lead to
a mixing of $\left| \delta \vartheta \right| $ and $\left| \delta I\right| $%
. The term $\mathrm{d}\omega /\mathrm{d}I$ will lead, in turn, to further
growth of $\left| \delta \vartheta \right| .$ As this process repeats
itself, both $\left| \delta I\right| $ and $\left| \delta \vartheta \right| $
are expected to grow rapidly. Thus ray instability is expected to be
significantly enhanced when $\left| \mathrm{d}\omega /\mathrm{d}I\right| $
is large.

\subsection{\textsf{KAM theory}}

A (background) Hamiltonian is said to be\textit{\ \textbf{nondegenerate} }if
$\mathrm{d}\omega /\mathrm{d}I\neq 0$, i.e. if the frequency varies from
torus to torus, which is a condition for nonlinearity of the system. An
important result for near-integrable nondegenerate Hamiltonian systems is
the celebrated\textit{\ \textbf{Kolmogorov-Arnold-Moser }}\textbf{(}\textit{%
\textbf{KAM}}\textbf{)}\textit{\textbf{\ theorem} }on the stability of
periodic solutions; cf. e.g. Refs.
%TCIMACRO{\TeXButton{Arnold-89,Tabor-89}{\onlinecite{Arnold-89,Tabor-89}}}%
%BeginExpansion
\onlinecite{Arnold-89,Tabor-89}%
%EndExpansion
. This theorem states that if $\delta \bar{h}$ is small enough, for most
initial conditions the motion remains periodic (i.e. confined to tori) and
the complement of the periodic motion (i.e. the chaotic motion) has a
measure that tends to zero as $\delta \bar{h}\rightarrow 0$. The KAM theorem
thus guarantees that periodic motion (i.e. the KAM tori) separate the
destroyed tori, leading to the notion of ``islands'' of (eternal) stability
immersed in a chaotic ``sea.''

The mechanism that produces the destruction of tori is\
trajectory--\allowbreak medium resonance; cf. e.g. Refs.
%TCIMACRO{\TeXButton{Arnold-89,Tabor-89}{\onlinecite{Arnold-89,Tabor-89}}}%
%BeginExpansion
\onlinecite{Arnold-89,Tabor-89}%
%EndExpansion
. For example, assume the perturbation Hamiltonian $\delta \bar{h}$ to be
periodic in range with frequency $\Omega $. Then it can be represented in
Fourier series $\delta \bar{h}=\func{Re}\sum_{m,n}A_{mn}(I)\exp \mathrm{i}%
(m\vartheta -n\Omega r)$. For KAM or\ \textit{nonresonant} tori, $m\omega
+n\Omega \neq 0$ for all integers $n,m$ and the motion is periodic. In
contrast, those tori that satisfy $m\omega +n\Omega =0$ for some integers $%
n,m$ are said to be in \textit{resonance} and chaotic motion develops. If
several tori are captured into resonance, then the character of the chaotic
motion will depend on whether these resonances overlap or not. For instance,
consider two resonances centered at $I_{1}$ and $I_{2}.$ The widths of these
resonances can be estimated as $\Delta I_{i}=4\sqrt{A_{i}/\left| \mathrm{d}%
\omega _{i}/\mathrm{d}I\right| }$, where $A_{i}$ is the amplitude of the
resonant term in the above expansion. Define then $\Delta I=\Delta
I_{1}+\Delta I_{2}.$ A benign form of chaos\ is present when these
resonances are isolated, i.e. $\Delta I<|I_{1}-I_{2}|$;\ strong chaos,\ in
contrast, emerges when these resonances overlap, i.e. $\Delta
I>|I_{1}-I_{2}| $. The last criterion, due to Chirikov \cite{Chirikov-79},
gives thus a quantitative estimate of the size of the region of phase space
occupied by chaotic trajectories.

The focus in KAM theory on the role of individual ray--medium resonances
might seem to be at odds conceptually with the heuristic argument given at
the end of the preceding subsection. There it was argued that the second
derivatives of the perturbation to the environment could be treated as
random variables. Ref.
%TCIMACRO{\TeXButton{Brown-98}{\onlinecite{Brown-98}} }%
%BeginExpansion
\onlinecite{Brown-98}
%EndExpansion
partially bridges this conceptual gap by showing that the KAM theorem can be
applied to problems for which the perturbation consists of a superposition
of an arbitrarily large finite number of frequencies. Chirikov's criterion
is still applicable, but its evaluation seems feasible only if the number of
frequencies that comprises the perturbation is very small.
\begin{figure}[tbp]
\centerline{\includegraphics[width=4cm]{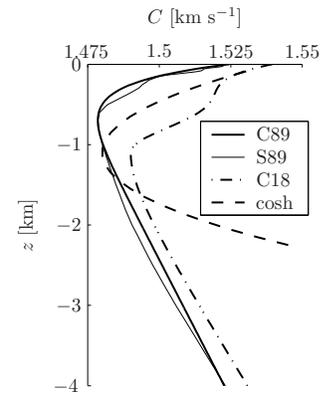}}
\caption{Background sound speed profiles used in the numerical work
presented in this paper.}
\label{SndProf}
\end{figure}

\section{\textsf{Numerical simulations}\label{NumSim}}

In this section numerical simulations are presented which were chosen to
demonstrate the importance of the background sound speed structure on ray
stability. We consider first deep-water conditions and upward-refracting
conditions afterward.

\subsection{\textsf{Deep-water conditions}}

Four different background profiles are studied here (cf. Fig. \ref{SndProf}%
). These are designated S89, C89, C18, and cosh. The S89 profile corresponds
to a range average of the Slice89 sound speed observations. The C89 profile
is a\textit{\ }canonical\textit{\ }profile \cite{Munk-74} with parameters
chosen to approximately match S89's sound channel axis depth, axial sound
speed, and surface sound speed. The C18 profile can be regarded as an
idealized model of the sound speed structure in the North Atlantic, whose
upper ocean structure is associated with the $18\unit{%
%TCIMACRO{\U{2103}}%
%BeginExpansion
{}^{\circ}{\rm C}%
%EndExpansion
}$ water mass \cite{Brown-etal-03}. Finally, the cosh profile, with
cosh-dependence on depth relative to the axis, was chosen because it has the
special property $\mathrm{d}\omega /\mathrm{d}I=0$ for all $I.$

The same internal-wave-induced sound speed perturbation field is
superimposed on all four background profiles. This field is assumed to
satisfy the relationship
\begin{equation}
\delta c/c=\mu N^{2}\zeta ,  \label{dc}
\end{equation}%
where $\zeta (z;r)$ is the internal-wave-induced vertical displacement of a
water parcel and $N(z;r)$ is the Brunt--V\"{a}is\"{a}l\"{a} frequency.
Relationship (\ref{dc}) with $\mu =1.25\;\mathrm{\unit{s}}^{2}\mathrm{\unit{m%
}}^{-1}$ was found to give a good fit to AET hydrographic measurements \cite%
{Wolfson-Spiesberg-99}. Our simulated internal-wave-induced sound speed
perturbations are similar to those used in Refs.
%TCIMACRO{
%\TeXButton{Brown-etal-03,Beron-etal-02}{\onlinecite{Brown-etal-03,Beron-etal-02}}}%
%BeginExpansion
\onlinecite{Brown-etal-03,Beron-etal-02}%
%EndExpansion
; these are based on Eq. (\ref{dc}) and make use of the $N$ profile
estimated from measurements made during the AET experiment. The statistics
of $\zeta $ are assumed to be described by the empirical Garrett--Munk
spectrum \cite{Munk-81}. The vertical displacement $\zeta $ is computed
using Eq. (19) of Ref.
%TCIMACRO{\TeXButton{Colosi-Brown-98}{\onlinecite{Colosi-Brown-98}} }%
%BeginExpansion
\onlinecite{Colosi-Brown-98}
%EndExpansion
with the variable $x$ replaced by $r$ and $y=0=t$. Physically this
corresponds to a frozen vertical slice of the internal wave field that
includes the influence of transversely propagating internal wave modes. A
Fourier method is used to numerically generate the sound speed perturbation
fields. A mode number cutoff of $30$ and a horizontal wavenumber cutoff of $%
2\pi $ $\unit{km}^{-1}$ were used in our simulations. The internal wave
strength parameter was taken to be the nominal Garrett--Munk value.
\begin{figure}[tbp]
\centerline{\includegraphics[width=8cm]{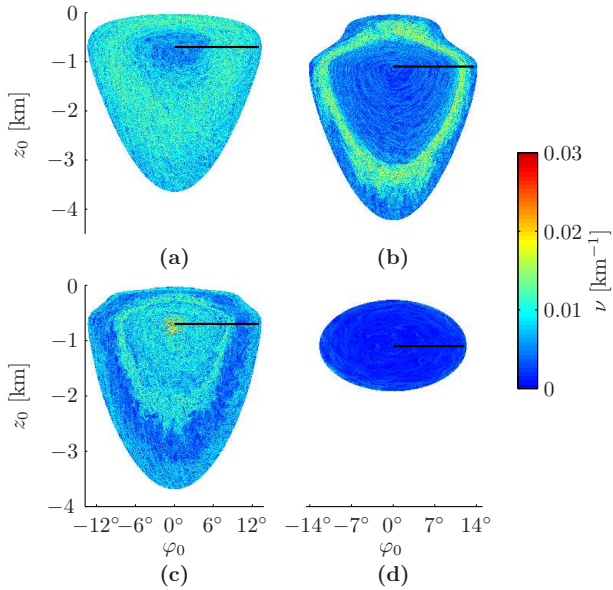}}
\caption{Stability exponents (finite range estimates of Lyapunov exponents)
as a function of the initial depth and launch angle, for ray simulations in
the (a) C89, (b) C18, (c) S89, and (d) cosh waveguides (cf. Fig. 1) with
internal-wave-induced perturbations superimposed.}
\label{NuPhaseSpace}
\end{figure}

Fig. \ref{NuPhaseSpace} shows the stability exponent, $\nu $ (a finite-range
estimate of the Lyapunov exponent; cf. appendix A), as a function of initial
ray position in phase space for each of the four environments considered.
This figure provides a general picture of the ray motion stability character
in each of the waveguides. In the C89 waveguide, ray trajectories with small
(resp., large) unperturbed ray axial angles have small (resp., large)
associated stability exponents. This trend is reversed in the S89 waveguide.
The disparity in the stability properties of the ray motion in these
waveguides contrasts with the close similarity of the corresponding
background sound speed profiles. In the C18 waveguide ray trajectories have
in general relatively small associated stability exponents, except in a
narrow band of initial actions (or unperturbed axial angles) where the
exponents are large and the ray motion more unstable. In opposition to the
other waveguides, in the cosh profile the stability exponents are very small
everywhere in phase space and the ray motion is remarkably stable.

\begin{figure}[tbp]
\centerline{\includegraphics[width=8cm]{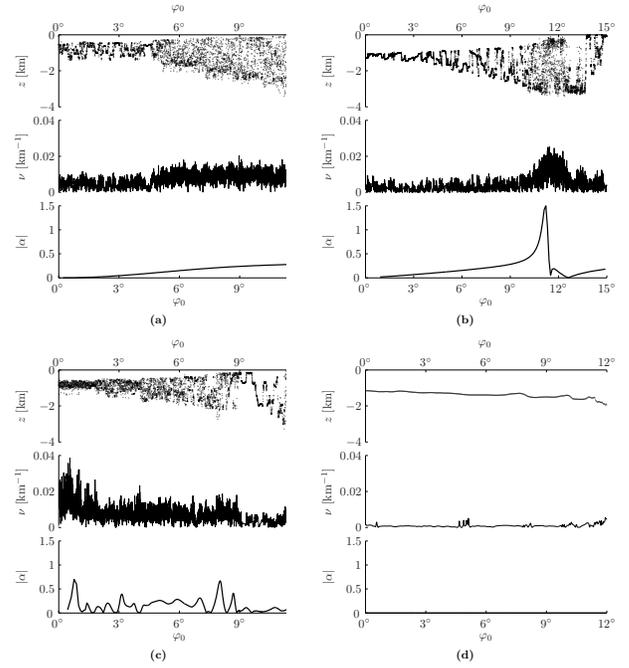}}
\caption{Each panel shows ray final depth (upper-plot), stability exponent
(middle-plot), and stability parameter (lower-plot) at $1000$ $\unit{km}$
range as a function of ray launch angle for a source on the sound channel
axis. The sound speed structures are identical to those used in the panels
of Fig. 2 with the same labels.}
\label{AlphaFig}
\end{figure}

In each panel of Fig. \ref{NuPhaseSpace} the horizontal line shown
corresponds to a fan of rays, launched on the sound channel axis, with
positive angles. For these rays, in each environment ray depth at a range of
$1000$ $\unit{km}$ and stability exponent are plotted as a function of
launch angle, $\varphi _{0},$ in Fig. \ref{AlphaFig}. In that figure bands
of regular trajectories appear as a smooth slowly-varying curve $z(\varphi
_{0})$ with small associated values of $\nu $; bands of chaotic rays appear
as a highly structured $z(\varphi _{0})$ curve with large associated values
of $\nu $. Also shown in Fig. \ref{AlphaFig} is a plot of the stability
parameter $\alpha $ defined in Eq. (\ref{Alpha}) vs. $\varphi _{0}$ in each
environment. (The action $I$ is a monotonically increasing function of the
axial ray angle which, for rays starting at the axis, coincides with $%
\varphi _{0};$ thus replacing $I$ by $\varphi _{0}$ represents a simple
stretching of the abscissa.) Notice in the $z$ vs. $\varphi _{0}$ plots the
seemingly unstructured (resp., ordered) distribution of points associated
with those angular bands where $\nu $ is large (resp., small). Notice also
that $\nu $ (or, equivalently, the irregularity of $z$ as a function of $%
\varphi _{0}$) appears to increase with increasing $\left| \alpha \right| .$
In particular, note that $\alpha =0$ (since $\mathrm{d}\omega /\mathrm{d}I=0$%
) for all $\varphi _{0}$ in the cosh profile. In this case ray final depth
varies very slowly with $\varphi _{0},$ and the stability exponents are very
small.

The numerical results presented in Figs. \ref{NuPhaseSpace} and \ref%
{AlphaFig} suggest that ray stability is strongly influenced by the
background sound speed structure and that ray instability increases with
increasing $\left| \alpha \right| .$ These results are consistent with the
heuristic argument given a the end of Sec. \ref{TheoBack} based on the
action--angle form of the ray variational equations; there it was argued
that ray instability should increase with increasing $\left| \mathrm{d}%
\omega /\mathrm{d}I\right| .$

\subsection{\textsf{Upward-refracting conditions}}

The validity of the argument given in Sec. \ref{TheoBack} is not limited to
deep ocean conditions, strongly suggesting that the result should be more
generally valid. We now present numerical results that support this
expectation.
\begin{figure}[t]
\centerline{\includegraphics[width=4cm]{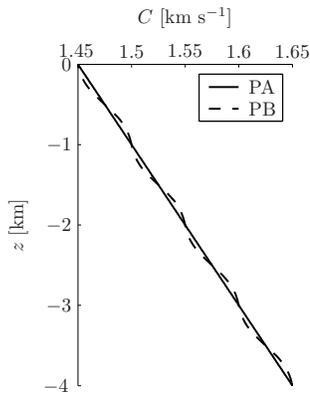}}
\caption{Background sound speed profiles used to construct the curves in
Fig. 5.}
\label{SndProfPolar}
\end{figure}

\begin{figure}[b]
\centerline{\includegraphics[width=8cm,clip=]{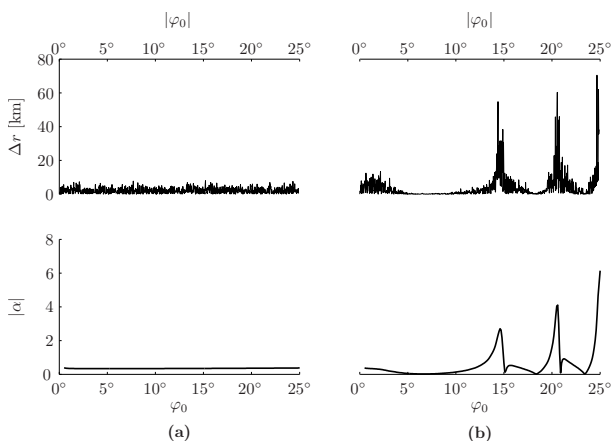}}
\caption{Upper panels: absolute value of the difference between perturbed
(rough surface) and unperturbed (flat surface) ray range after 20 surface
reflections as a function of initial ray angle at the surface for sound
speed profiles PA (a) and PB (b) in Fig. 4. Lower panels: stability
parameter as a function of ray angle in each of the two environments.}
\label{AlphaPolar}
\end{figure}

Figure \ref{SndProfPolar} shows two upward-refracting sound speed profiles.
Figure \ref{AlphaPolar} shows $\alpha $ in the same environments and the
difference in range, $\Delta r,$ between perturbed (rough surface) and
unperturbed (flat surface) rays as a function of launch angle at the surface
after making 21 loops, which corresponds to 20 surface reflections. [In this
type of environment the definition of $I$ in Eq. (\ref{I}) is unchanged
except that the upper integration limit is $z=0$ for all rays.] Rough
surface scattering was treated using a frozen simulated surface gravity
wavefield with a $k^{-7/2}$ surface elevation wavenumber spectrum with $0.02$
\textrm{rad m}$^{-1}\leq k\leq 0.16$ $\mathrm{rad}$ \textrm{m}$^{-1}$, $%
\Delta k=10^{-3}$ \textrm{rad m}$^{-1},$ and rms slope of $4\times 10^{-3}.$
To treat specular ray reflections from this surface, the surface boundary
condition was linearized; the surface elevation was neglected, but the
nonzero slope was not approximated.

Figure \ref{AlphaPolar} shows clearly that ray stability, as measured by $%
\Delta r$, is controlled almost entirely by the background sound speed
structure via $\alpha $, rather than details of the rough surface.

\section{\textsf{Shear-induced ray instability}\label{ShearIns}}

In this section we present additional numerical simulations that give
insight into the mechanism by which $\alpha $ influences ray instability.
For simplicity we assume deep water conditions here. The arguments presented
here make use of the well-known (cf. e.g. Refs.
%TCIMACRO{\TeXButton{Arnold-89,Tabor-89}{\onlinecite{Arnold-89,Tabor-89}}}%
%BeginExpansion
\onlinecite{Arnold-89,Tabor-89}%
%EndExpansion
) analogy between ray motion, defined by Eqs. (\ref{RayEqn}) and particle
motion in an incompressible two-dimensional fluid.

\begin{figure}[tbp]
\centerline {\includegraphics[width=6cm]{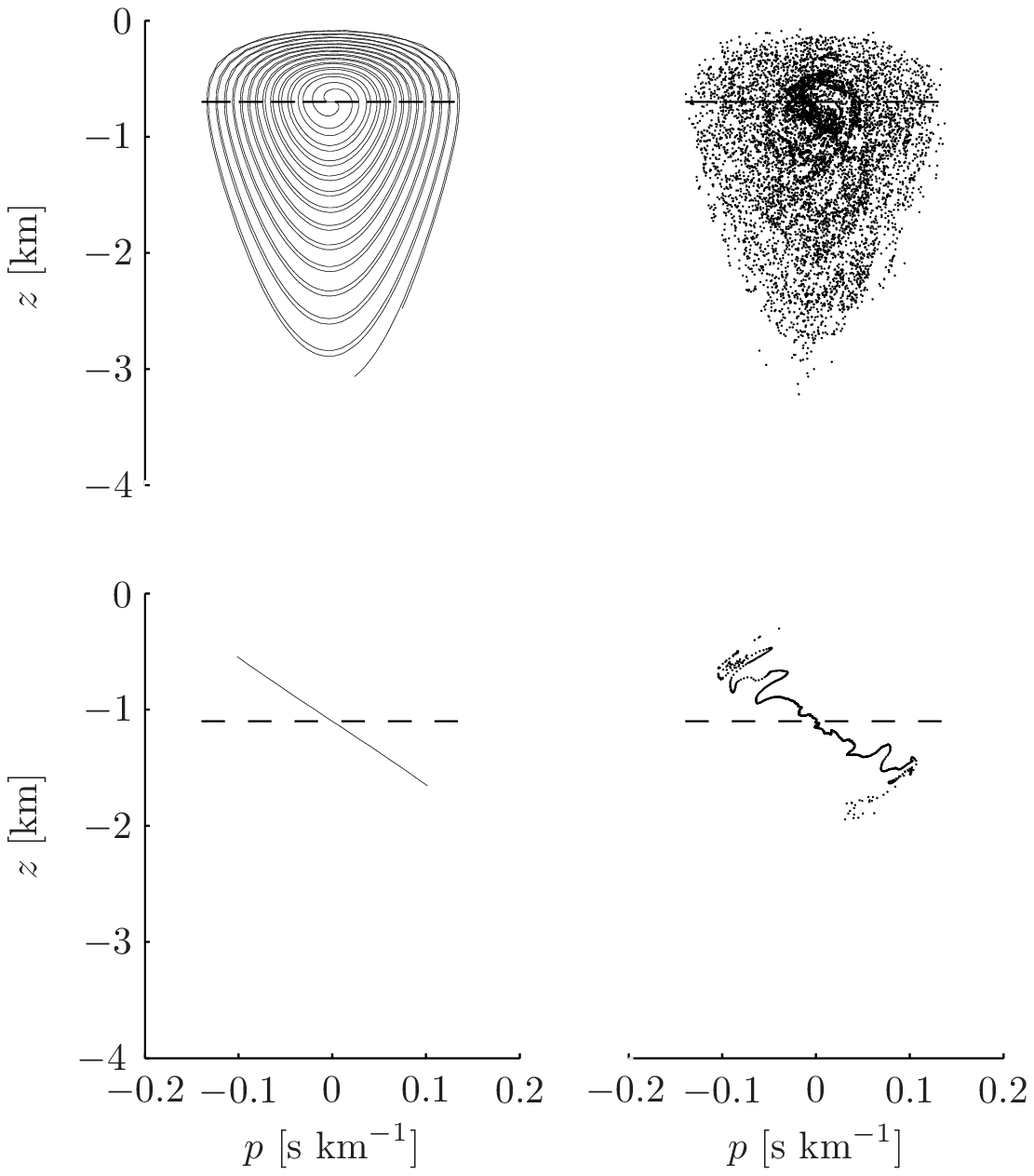}}
\caption{Evolution of a portion of Lagrangian manifold in the C89 (upper
panels) and cosh (lower panels) waveguides with (right panels) and without
(left panels) internal-wave-induced perturbations. Dashed and solid/dotted
curves show the manifold at $r=0$ and $r=1000$ $\unit{km}$, respectively.}
\label{MunkVsChLM}
\end{figure}

The left panels of Fig. \ref{MunkVsChLM} show the range evolution of a
segment of a\textbf{\ \textit{Lagrangian manifold} }(a smooth curve in phase
space) in the unperturbed C89 (top) and cosh (bottom) waveguides. The
segment is depicted in black at $r=0$ and in red at $r=1000$ $\unit{km}$ in
both waveguides. As a consequence of Liouville's theorem the segment cannot
break or intersect itself but it can increase in complexity as range
increases. In the unperturbed case, since the motion is integrable (i.e.
each point of the segment preserves its initial $I$), the length of the
segment can grow in range, at most, following a power law. The cosh profile
has a special property. In that profile the manifold segment just rotates
counterclockwise at a constant rate $\omega $. The reason for the difference
in behavior is that in the C89 waveguide $\omega $ varies with $I$, whereas
in the cosh profile $\omega $ does not vary with $I$. The monotonic decay of
$\omega $ as a function of $I$ in the C89 waveguide induces a\textit{\ ``}%
shear'' in phase space which causes the outermost points of the segment to
rotate more slowly than the innermost ones and, hence, causes the segment to
spiral. The ray motion in phase space associated with the unperturbed C89
waveguide can thus be regraded as analogous to that of ideal fluid particles
passively advected by a stationary planar circular flow with radial shear.
In the cosh profile there is no shear. In polar coordinates \textit{\textbf{%
radial shear} }can be defined as
\begin{equation}
\rho \frac{\partial }{\partial \rho }\left( \frac{u_{\theta }}{\rho }\right)
,  \label{shear}
\end{equation}%
where $\rho $ is the radial coordinate and $u_{\theta }$ is the $\theta $%
-component of the velocity field. (More correctly, this quantity is, apart
from a factor of $2$, the $\rho \theta $-component of the strain-rate tensor
for planar circular flow; cf. e.g. Ref.
%TCIMACRO{\TeXButton{Batchelor-64}{\onlinecite{Batchelor-64}}}%
%BeginExpansion
\onlinecite{Batchelor-64}%
%EndExpansion
.) The connection with ray motion in phase space can be accomplished by
identifying $I$ with $\rho $ and $\omega I$ with $u_{\theta }$. The
replacements $\rho \mapsto I$ and $u_{\theta }\mapsto \omega I$ in Eq. (\ref%
{shear}) thus give the analogous expression $I\mathrm{d}\omega /\mathrm{d}I$
for the shear in phase space. Notice that this expression is (apart from the
$\omega ^{-1}$-factor) the stability parameter $\alpha $. We have chosen to
include the $\omega ^{-1}$-factor in the definition of $\alpha $ because of
precedent \cite{Zaslavsky-98} and because it is convenient to make $\alpha $
dimensionless.

The right panels of Fig. \ref{MunkVsChLM} show the evolution of the
Lagrangian manifold segment in the same waveguides as those used to produce
the left panels but with a superimposed perturbation induced by internal
wave fluctuations. Notice the highly complicated structure of the Lagrangian
manifold in the perturbed C89 waveguide as compared to that in the
unperturbed one. (Note that the fan of rays used to produce the figure is
far too sparse to resolve what should be an unbroken smooth curve which does
not intersect itself.) In contrast, observe that in the cosh environment the
sound speed perturbation has only a very minor effect on the evolution of
the Lagrangian manifold.

\begin{figure}[t]
\centering\subfigure[]{\epsfig{file=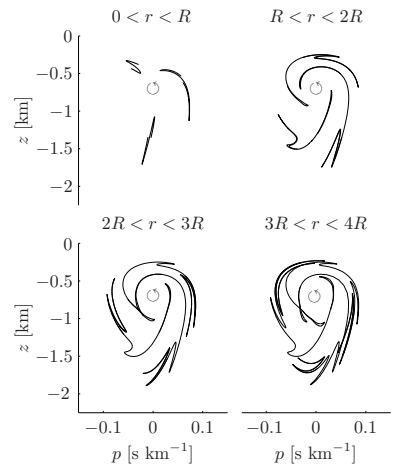,width=5cm,clip=}}\qquad%
\subfigure[]{\epsfig{file=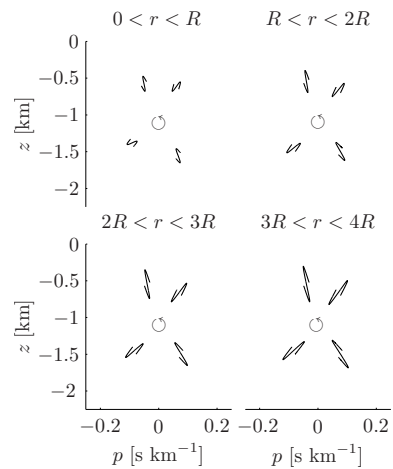,width=5cm,clip=}}
%\centerline{\includegraphics[width=9cm]{MunkVsChLMApex.eps}}
\caption{Evolution of a segment of Lagrangian manifold in the C89 (a) and
cosh (b) waveguides under the influence of an idealized range-dependent
perturbation. The perturbations are is assumed to be in the form of a series
of ``kicks'' that produce a sinusoidal deformation to the Lagrangian
manifold at each upper turning depth. The figure shows a portion of
Lagrangian manifold, originally located on a torus with frequency $\protect%
\omega$($=2\protect\pi /R$)$,$ immediately after experiencing a
kick at range $r=(k-1)R$, $k=1,\cdots ,4,$ and before receiving
the next one at $r=kR $, as well as at several intermediate
stages.} \label{MunkVsChLMApex}
\end{figure}

Perturbations to steeper rays caused by internal-wave-induced sound speed
perturbations in deep ocean, including those in our simulations, are
significant only near the ray's upper turning depth. This observation
motivates a simple model that gives insight into the mechanism by which $%
\alpha $ is linked to ray stability. In the model, each portion of a
Lagrangian manifold acquires a sinusoidal ``wrinkle'' at each upper turning
point, but is otherwise unaffected by the sound speed perturbations. The
evolution of small segments of a Lagrangian manifold using such a model, in
both the C89 and cosh waveguides, is shown in Fig. \ref{MunkVsChLMApex}.
Notice the rapid growth in complexity of the segment as the range increases
in the C89 waveguide. After acquiring a wrinkle, the segment stretches and
folds as a result of the radial shear in phase space. As additional wrinkles
are acquired this process is repeated successively in range, making the
shape of the Lagrangian manifold segment even more complex. In opposition to
this situation, observe the simplicity of the segment's shape in the cosh
waveguide as range increases. In this case, after acquiring a wrinkle, the
distorted segment rotates counterclockwise, without the additional influence
of shear-induced stretching, at a constant frequency $\omega $.

Each time a perturbation is introduced, the action $I$ changes by the amount
$\delta I,$ say, which we assume to be of the same order as the
perturbation. As a consequence, to lowest-order $R$($=2\pi /\omega $)
experiences the change%
\[
R\mapsto \left( 1-\alpha \,\delta I/I\right) R.
\]%
The perturbation to the range of a ray double loop $R$ depends on both the
perturbation \textit{and} the properties of the background sound speed
structure. Under the change $I\mapsto I+\delta I,$ a sufficient condition
for $R$ to remain invariant at lowest-order is $\alpha =0.$ This provides an
explanation for the remarkable stability of the cosh waveguide. That is,
when $\alpha =0$ the ray motion remains periodic at lowest-order---no matter
the complexity of the perturbation mounted on the waveguide. To
lowest-order, a nonvanishing shear $(\alpha \neq 0)$ appears as a necessary
condition to sustain the successive stretching and folding of the Lagrangian
manifold after it gets distorted by the perturbation. (Of course, chaotic
motion is still possible when $\alpha =0$ provided that the perturbation
strength is sufficiently large.) It is thus expected that where $|\alpha |$
is small (resp., large) there will be less (resp., more) sensitivity to
initial conditions and, hence, the motion be more regular (resp., chaotic).
Support for this conjecture is given in the numerical simulations presented
in this paper.

\section{\textsf{Summary and discussion}\label{SumDis}}

In this paper we have considered ray motion in environments consisting of a
range-independent background sound speed profile on which a weak
range-dependent perturbation is superimposed. The results presented show
that ray stability is strongly influenced by the background sound speed
structure; ray instability was shown to increase with increasing magnitude
of $\alpha (I)=(I/\omega )\mathrm{d}\omega /\mathrm{d}I,$ where $2\pi
/\omega (I)$ is the range of a ray double loop and $I$ is the ray action
variable. This conclusion is based largely on numerical simulations, but the
simulations were shown to support a simple heuristic argument based on the
action--angle form of the ray variational equations. The mechanism by which $%
\alpha $ controls ray instability was shown to be shear-induced enhancement
of perturbations caused by the sound speed perturbation term. The importance
of $\alpha $ was illustrated with numerical simulations of ray motion in
deep ocean environments including internal-wave-induced scattering, and in
upward-refracting environments including rough surface scattering. So far as
we are aware, this conclusion is consistent with all of the numerical
results presented earlier \cite%
{Duda-Bowlin-94,Simmen-Flatte-Yu-Wang-97,Smirnov-Virovlyansky-Zaslavsky-01}.
Ref.
%TCIMACRO{
%\TeXButton{Smirnov-Virovlyansky-Zaslavsky-01}{\onlinecite{Smirnov-Virovlyansky-Zaslavsky-01}} }%
%BeginExpansion
\onlinecite{Smirnov-Virovlyansky-Zaslavsky-01}
%EndExpansion
relied heavily on results that follow from a dynamical systems viewpoint.

The connection between our work and results relating to dynamical systems
deserves further comment. Recall that the condition $\mathrm{d}\omega /%
\mathrm{d}I\neq 0$ (the nondegeneracy condition) must be satisfied for the
KAM theorem to apply, and that this result guarantees that some rays are
nonchaotic provided the strength of the range-dependent perturbation is
sufficiently weak. This theorem might seem to conflict with our assertion
that ray instability increases with increasing $\left| \alpha \right| .$
This apparent conflict can be resolved by interpreting our statement as a
statement of what happens for most rays. That is, for most rays stability
exponents (finite range estimates of Lyapunov exponents) increase as $\left|
\alpha \right| $ increases. This does not rule out the possibility that for
a fixed $\mathrm{d}\omega /\mathrm{d}I\neq 0$ some rays will be nonchaotic.

A seemingly more troublesome conflict between our simulations and KAM theory
follows from the result, noted earlier, that each isolated resonance has a
width proportional to $|\mathrm{d}\omega /\mathrm{d}I|^{-1/2}$. Doesn't this
imply that rays should become increasingly chaotic as $|\mathrm{d}\omega /%
\mathrm{d}I|$ decreases? The answer, we believe, is no. To understand why,
consider rays in a band $I_{0}\leq I\leq I_{1}$ over which $\mathrm{d}\omega
/\mathrm{d}I$ is approximately constant, $\mathrm{d}\omega /\mathrm{d}%
I=(\omega _{1}-\omega _{0})/(I_{1}-I_{0})$. Within this band resonances are
excited at selected values of $\omega $. The number of resonances excited is
approximately proportional to $|\omega _{1}-\omega _{0}|$, which, in turn,
is proportional to $|\mathrm{d}\omega /\mathrm{d}I|$. The ``degree of
chaos'' should be proportional to the product of the number of resonances
excited and the width of individual resonances. This product scales like $|%
\mathrm{d}\omega /\mathrm{d}I|^{1/2}$; this suggests that rays should become
increasingly chaotic as $|\mathrm{d}\omega /\mathrm{d}I|$ increases,
consistent with our simulations. Neither the KAM theorem nor the resonance
width estimate estimate applies in the limit $\mathrm{d}\omega /\mathrm{d}%
I=0 $. Behavior in that limit is likely problem-dependent (e.g. Refs.
%TCIMACRO{
%\TeXButton{Chernikov-et-al-87,Chernikov-Sagdeev-Zaslavsky-88}{\onlinecite{Chernikov-et-al-87,Chernikov-Sagdeev-Zaslavsky-88}}}%
%BeginExpansion
\onlinecite{Chernikov-et-al-87,Chernikov-Sagdeev-Zaslavsky-88}%
%EndExpansion
). Our simulations---which are probably representative of problems
characterized by an inhomogeneous background and a weak perturbation with a
broad spectrum---suggest that ray motion in this limit is very stable.

In this paper we have addressed the issue of ray stability in physical space
or phase space; we have not addressed the related problem of travel time
stability. The latter problem is more difficult inasmuch as that problem
involves, in addition to the (one-way) ray equations (\ref{RayEqn}), a third
equation $\mathrm{d}T/\mathrm{d}r=L$ and imposition of an eigenray
constraint. Here, $L=p\mathrm{d}z/\mathrm{d}r-h(p,z;r)$ using standard
variables, or $L=I\mathrm{d}\vartheta /\mathrm{d}r-\bar{h}(I,\vartheta ;r)$
using action--angle variables. We have seen some numerical evidence that ray
instability in phase space is linked to large time spreads, but this
connection is currently not fully understood.

An advantage of our use of the action--angle formalism is that essentially
the same results apply if the assumption of a background range-independent
sound speed profile, i.e. $C=C(z),$ is relaxed to allow for a slowly-varying
(in range) background sound speed structure, i.e. $C=C(z;\varepsilon r).$
(Here, $\varepsilon $ scales like the ratio of a typical correlation length
of the range-dependence to a typical ray double loop range.) In the latter
case the action is not an exact ray invariant but it is an adiabatic
invariant, i.e. $\mathrm{d}I/\mathrm{d}r=O(\varepsilon ^{2}).$ Thus, correct
to $O(\varepsilon )$ the latter problem can be treated as being identical to
the former one. Consequently, the problem that we have treated here also
applies to slowly-varying background environments.

\section*{\textsf{Acknowledgments}}

We thank J. Colosi, S. Tomsovic, A. Virovlyansky, M. Wolfson, and
G. Zaslavsky for the benefit of discussions on ray chaos. The
comments of anonymous reviewers lead to improvements in the paper.
This research was supported by Code 321OA of the Office of Naval
Research.

\section*{\textsf{APPENDIX: STABILITY EXPONENTS}}

Liouville's theorem (cf. e.g. Refs.
%TCIMACRO{\TeXButton{Arnold-89,Tabor-89}{\onlinecite{Arnold-89,Tabor-89}}}%
%BeginExpansion
\onlinecite{Arnold-89,Tabor-89}%
%EndExpansion
) guarantees that areas in the present two-dimensional phase space are
preserved, leading to $\det \mathsf{Q}=1$ as a corollary. Consequently, $%
2\nu _{\pm }^{\mathsf{Q}}=\limfunc{trace}\mathsf{Q}\pm \sqrt{\limfunc{trace}%
^{2}\mathsf{Q}-1}$. (It can be shown that $\nu _{\pm }^{\mathsf{Q}}\sim \limfunc{trace}^{\pm 1}%
\mathsf{Q}$ as $r\rightarrow \infty $; accordingly, Eq.
(\ref{LyaExp}) can
be replaced by the equivalent expression $\nu =$ $\lim\nolimits_{r%
\rightarrow \infty }$ $r^{-1}\ln \left| \limfunc{trace}\mathsf{Q}\right| $
as in Ref. 20.) \nocite{Wolfson-Tomsovic-01} The condition $\det \mathsf{Q}%
=1 $, however, is difficult to fulfill in highly chaotic flows due to the
limitation of machine numerical precision. More precisely, an initial area
in phase space tends to align along the one-dimensional perturbation
subspace spanned by the eigenvector of $\mathsf{Q}$ associated with the
largest Lyapunov exponent, making the elements of $\mathsf{Q}$
ill-conditioned. This implies that neither $\det \mathsf{Q}$ nor $\limfunc{%
trace}\mathsf{Q}$---and, hence, neither $\nu ^{\mathsf{Q}}$ nor $\nu
_{\infty }$---can be computed reliably at long range. The common approach to
overcome this problem involves successive normalizations of the elements of $%
\mathsf{Q},$ while integrating simultaneously (\ref{RayEqn}) and (\ref%
{RayVar}), after a fixed number of range steps \cite{Parker-Chua-89}. An
alternative strategy, which does not involve normalizations, has been taken
here.

Let $\mathsf{A}(r)$ be the bounded-element matrix such that%
\begin{equation}
\mathsf{Q}=\mathsf{A}\,\mathrm{e}^{\lambda r},  \label{Q}
\end{equation}%
which implies%
\[
\nu ^{\mathsf{Q}}=\nu ^{\mathsf{A}}\mathrm{e}^{\lambda r}.
\]%
Here, $\lambda $ is a guessed value of $\nu _{\infty }.$ The reason for
introducing the decomposition (\ref{Q}) is that if $\lambda $ is close to $%
\nu _{\infty },$ the matrix $\mathsf{A}$ will remain well conditioned at
ranges long past those at which $\mathsf{Q}$ becomes poorly conditioned.
This leads, in turn, to significantly improved numerical stability. Eq. (\ref%
{RayVar}) leads to the\textit{\ \textbf{modified variational equations}\ }%
\begin{equation}
\frac{\mathrm{d}\mathsf{A}}{\mathrm{d}r}=\left( \mathsf{J}-\lambda \mathsf{I}%
\right) \mathsf{A},\quad \mathsf{A}(0)=\mathsf{I}.  \label{EqnForA}
\end{equation}%
Notice that $\det \mathsf{Q}=\det \mathsf{A\,}\mathrm{\exp }\lambda r=1$
and, hence, $\det \mathsf{A}\sim 0\;$as\ $r\rightarrow \infty .$ Eq. (\ref%
{EqnForA}) has been integrated in this paper after choosing $\lambda $, in
order to compute a finite range estimate of $\nu _{\infty }$, which we call a%
\textit{\ \textbf{stability exponent} }and denote by $\nu .$

\end{document}